\documentclass[aps,prl,twocolumn,showpacs,superscriptaddress]{revtex4}

\usepackage{graphicx,amsmath}% Include figure files

\newcommand{\be}{\begin{equation}}
\newcommand{\ee}{\end{equation}}
\newcommand{\bea}{\begin{eqnarray}}
\newcommand{\eea}{\end{eqnarray}}

\begin{document}

\title{Phenomenological model for two gap states 
in underdoped high-temperature superconductors
and short-range antiferromagnetic correlation effect}
\author{T. Morinari}
\email[]{morinari@yukawa.kyoto-u.ac.jp}
\affiliation{Yukawa Institute for Theoretical Physics, Kyoto
University, Kyoto 606-8502, Japan}
\date{\today}
\begin{abstract}
Assuming antiferromagnetic orbital correlations to 
model the pseudogap state in the underdoped high-temperature superconductors,
we study how this correlation is distinguished from the 
d-wave superconductivity correlation with including the finite-range
antiferromagnetic correlation effect.
In spite of the fact that both correlations have the same d-wave symmetry,
the contributions from each correlation is clearly distinguished
in the spectral weight and the density of states.
\end{abstract}

\pacs{
71.18.+y, % Fermi surface: calculations and measurements; effective mass, g factor
74.20.-z, % Theories and models of superconducting state
74.72.-h  % Cuprate superconductors (high-Tc and insulating parent compounds)
}
%71.10.-w, % Theories and models of many-electron systems
%71.10.Ay, % Fermi-liquid theory and other phenomenological models
%71.27.+a, % Strongly correlated electron systems; heavy fermions

\maketitle

%\section{Introduction}
In the high-temperature superconductors, the origin of the pseudogap state
is still in controversial.
Contrary to conventional superconductors, 
behaviors associated with opening some gap are observed in various
experiments above the transition temperature to the superconducting state,
$T_c$.\cite{Timusk1999}
Concerning the origin of the pseudogap,
mainly there are two pictures.
One is to assume that the pseudogap is associated 
with precursor of superconductivity.\cite{Emery1995}
The other is to assume another gap formation which is different from 
the superconducting gap.\cite{Chakravarty2001,WenLee1996,Varma1997,Vojta1999,Paramekanti2001}

Recent angle-resolved photoemission spectroscopy measurements
seem to support two gap scenario.\cite{Tanaka2006,Kondo2007}
In the normal state, the full Fermi surface is clearly observed.
Along this Fermi surface, the gap has been measured in the superconducting
state.
Around the nodal region, the gap is well fitted by
a d-wave superconducting gap.
As moving away from the node, the gap deviates from the line
expected from the simple d-wave superconducing gap.
The doping dependence of the gap around the node and 
that around the antinodal region are investigated.
It is found that the doping dependences of the gaps
in these different regions are qualitatively different.
The gap around the node increases as the hole doping concentration is increased
in the underdoped regime.
By contrast, the gap around the antinodal region decreases.
Furthermore, the gap around the antinode shows 
negligible change across $T_c$ whereas 
the gap around the node shows significant variation.
This result suggests that those gaps have different origins.
This is consistent with the recent Raman scattering experiments.\cite{Tacon2006}
In the Raman scattering experiments 
the nodal region and the antinodal region
are distinguished by symmetry.
From B$_{1g}$ and B$_{2g}$ spectra,
the Raman shifts show qualitatively different behaviors.
The shift associated with the node (B$_{2g}$)
increases as the doping concentration is increased
while the shift associated with the antinode (B$_{1g}$)
decreases.

The two gap picture is also supported by scanning tunneling
microscopy (STM).
About a decade ago, some STM measurements supported
that the pseudogap and the superconducting gap had the same origin
because the pseudogap seemed to smoothly evolve into the superconducting gap
as decreasing temperature.
However, in the recent STM measurements\cite{Boyer2007} 
it is found that there are two components:
One is inhomogeneous and the other is homogeneous.
The inhomogeneous component is associated with relatively high-energy 
features and it is found that that component does not show
clear temperature dependence around $T_c$.
Boyer {\it et al.} extracted the homogeneous component
from the raw data by taking a normalization using the spectra
taken above $T_c$.\cite{Boyer2007}
The gap in the homogeneous component clearly vanishes at $T_c$.
By contrast, the inhomogeneous component does not show qualitative
difference at $T_c$.
It is also found that the gap value estimated for the homogeneous
component is reduced from that estimated from the raw data.
It is known that the latter gap $\Delta_p$ 
is scaled by the pseudogap temperature, $T^*$ with 
$2\Delta_p \simeq 4.3 k_B T^*$.
This scaling relation is consistent with the orbital antiferromagnetic
correlation with d-wave symmetry. \cite{Won2007}

If both of superconductivity and the pseudogap 
are characterized by d-wave symmetry,
then the question is how we can distinguish 
these correlations in the physical quantities.
To answer this question,
we study a simple phenomenological model
for the two gap state for high-temperature superconductors.
We assume the d-wave BCS gap for superconductivity and
d-wave orbital antiferromagnetic correlation
\cite{AffleckMarston1988,Nersesyan1989,Chakravarty2001,Won2007}
to model the pseudogap.
Based on this model we compute the spectral weight 
along the underlying Fermi surface
and the density of states
with including the short-range antiferromagnetic correlation effect.
In spite of the fact that the two gaps have the same symmetry,
we find that these components are clearly distinguished 
in the Brillouin zone and in the density of states.

%The organization of the paper is as follows.
%In Sec.\ref{sec_formalism} we describe the model.
%We compute the spectral weight along the underlying Fermi surface.
%Section \ref{sec_} is devoted to summary and conclusion.

%\section{Formalism}
%\label{sec_formalism}
The model Hamiltonian is given by
\begin{widetext}
\bea
 H &=& \sum\limits_{k \in RBZ} {\left( {\begin{array}{*{20}c}
   {c_{k \uparrow }^\dag  } & {c_{ - k \downarrow } } & {c_{k + Q \uparrow }^\dag  } 
& {c_{ - k - Q \downarrow } }  \\
\end{array}} \right)} \left( {\begin{array}{*{20}c}
   {\varepsilon _k  - \mu } & {\Delta _k^{SC} } & {i\Delta _k } & 0  \\
   {\Delta _k^{SC} } & { - \varepsilon _k  + \mu } & 0 & {i\Delta _k }  \\
   { - i\Delta _k } & 0 & {\varepsilon _{k + Q}  - \mu } & { - \Delta _k^{SC} }  \\
   0 & { - i\Delta _k } & { - \Delta _k^{SC} } & { - \varepsilon _{k + Q}  + \mu }  \\
\end{array}} \right)\left( {\begin{array}{*{20}c}
   {c_{k \uparrow } }  \\
   {c_{ - k \downarrow }^\dag  }  \\
   {c_{k + Q \uparrow } }  \\
   {c_{ - k - Q \downarrow }^\dag  }  \\
\end{array}} \right) \nonumber \\
 & & + \sum\limits_{k \in RBZ} {\left[ {\left( {\varepsilon _k  - \mu } \right) 
+ \left( {\varepsilon _{k + Q}  - \mu } \right)} \right]}.
\label{eq_H}
\eea
\end{widetext}
where
$\varepsilon _k  =  - 2t_0 \left( {\cos k_x  + \cos k_y } \right) - 4t_1 \cos k_x \cos k_y  
- 2t_2 \left( {\cos 2k_x  + \cos 2k_y } \right)$,
$\Delta _k  = \frac{{\Delta _0 }}{2}\left( {\cos k_x  - \cos k_y } \right)$,
and 
$\Delta _k^{SC}  = \frac{{\Delta _0^{SC} }}{2}\left( {\cos k_x  - \cos k_y } \right)$.
The summation with respect to $k$ is taken over the reduced Brillouin zone,
$|k_x|<\pi$ and $|k_y|<\pi$.
The wave vector $Q$ is fixed to $Q=(\pi,\pi)$ for the pure mean field state.
In order to include the short-range antiferromagnetic correlation effect
approximately, we assume that $Q$ obeys a Lorentzian probability distribution
whose peak is at $(\pi,\pi)$ with broadening factor of $\xi_{AF}^{-1}$.
We follow the analysis described in Ref.\cite{Harrison2007} in investigating
the short-range antiferromagnetic correlation effect.

The energy dispersions are obtained by diagonalizing (\ref{eq_H}),
\be
E_k^{\left( { \pm , \pm } \right)}  =  
\pm \sqrt {\left( { \varepsilon _k^{\left(  +  \right)}  - \mu  
\pm \sqrt {\left( {\varepsilon _k^{\left(  -  \right)} } \right)^2  + \Delta _k^2 } } 
\right)^2  + \left( {\Delta _k^{SC} } \right)^2 }.
\ee
Here $\varepsilon _k^{\left(  +  \right)}  
= \frac{{\varepsilon _k  + \varepsilon _{k + Q} }}{2}$
and
$\varepsilon _k^{\left(  -  \right)}  
= \frac{{\varepsilon _k  - \varepsilon _{k + Q} }}{2}$.
For the calculation of the spectral weight, we compute
the Matsubara Green's function, 
$G_{k \uparrow } \left( \tau  \right) 
= - \left\langle {T_\tau  c_{k \uparrow } \left( \tau  \right)c_{k \uparrow }^\dag  
\left( 0 \right)} \right\rangle$.
Transforming to the Matsubara frequency from the imaginary time $\tau$, we have
\be
G_{k \uparrow } \left( {i\omega _n } \right) = 
%\sum\limits_{\alpha  = 1,2,3,4} 
\sum_{\alpha=1}^4
{\frac{{U_{k \uparrow ,\alpha } 
\left( {U^\dag  } \right)_{\alpha ,k \uparrow } }}{{i\omega _n  - E_\alpha  }}}.
\ee
The Unitary matrix $U_{k\uparrow,\alpha}$ is computed numerically diagonalizing 
Eq. (\ref{eq_H}).
The index $\alpha$ is for the energy bands.
We have four bands for $\Delta^{SC}_0 \neq 0$ and 
$\Delta_0 \neq 0$.

\begin{figure}
   \begin{center}
    \includegraphics[width=2.6in]{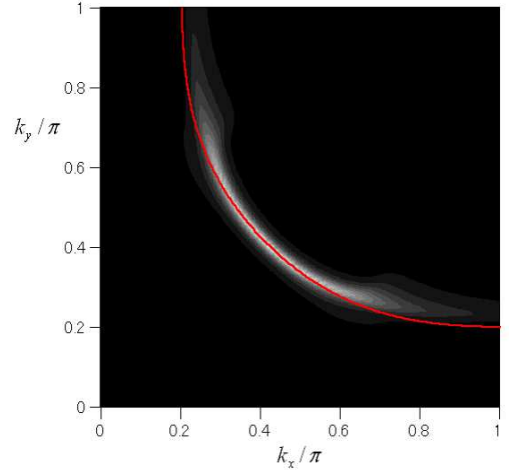}
   \end{center}
   \caption{ \label{fig_fs}
	Spectral weight at the Fermi energy averaged over probability distribution
	with respect to $Q$. 
	The antiferromagnetic correlation length is $\xi_{AF}=4$.
	The solid line represents
	the underlying Fermi surface with $\Delta^{SC}_0 = 0$ and 
	$\Delta_0=0$.
    }
 \end{figure}

In Fig.\ref{fig_fs}, the spectral weight at the Fermi energy is shown.
%=====================================================================
% [code] spectral weight: short_range_af/codes_sraf/*
%        underlying FS  : stm_theory/codes_stm/*
%        [Data file]    : *
%=====================================================================
The solid line represents the under-lying Fermi surface which 
is given by $\varepsilon_k = \mu$.
The chemical potential $\mu$ is determined from 
\be
x = 1 - \frac{1}{{2N_{s} }}\sum\limits_k {n_k },
\ee
with $n_k  = 2 + \left\langle {c_{k \uparrow }^\dag  c_{k \uparrow } } 
\right\rangle  - \left\langle {c_{ - k \downarrow } c_{ - k \downarrow }^\dag  } 
\right\rangle  + \left\langle {c_{k + Q \uparrow }^\dag  c_{k + Q \uparrow } } 
\right\rangle  - \left\langle {c_{ - k - Q \downarrow } c_{ - k - Q \downarrow }^\dag  } 
\right\rangle$,
$x$ the doped hole concentration and $N_{s}$ the number of the lattice sites.
The expectation values are calculated by diagonalizing (\ref{eq_H}).
We take the hopping parameters as $t_0=1$, $t_1=-0.25$, and $t_2=0.10$ through out 
the paper.

%%%%%%%%%%%%%%%%%%%%%%%%%%%%%%%%%%%%%%%%%%%%%%%%%%%%%%%%%%%%%%%%%%%%%%
% code: sraf_effect/codes_srafe/smh1_0.cc
% Data: Data0613_08/datd_xiAF_1p0 (xiAF=1)
% Data: Data0613_08/datd_xiAF_4p0 (xiAF=4)
Although we assume a finite antiferromagnetic correlation length,
the mean field calculation shows that $2\Delta_0/k_B T^* \simeq 4.8$
for the interaction $V=2$ and $\xi_{AF}=4$ on $50\times 50$ lattice.
Therefore, we may assume a finite value of $\Delta_0$
in spite of the lack of the orbital antiferromagnetic long range ordering.

\begin{figure}
   \begin{center}
    \includegraphics[width=3.4in]{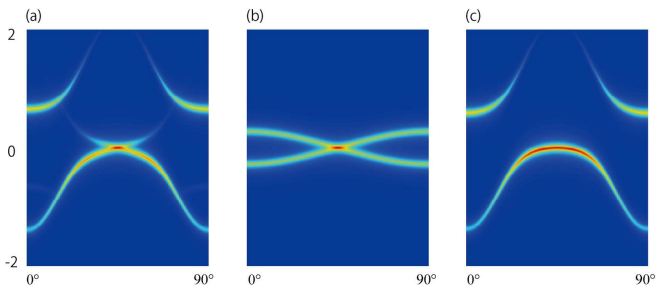}
   \end{center}
   \caption{ \label{fig_spA_along_fs}
	Spectral weight along the underlying Fermi surface.
	The vertical axis represents the energy and the horizontal axis
	represents the angle along the underlying Fermi surface:
	(a)$\Delta_0/t_0=1.0$ and $\Delta^{SC}_0=0.3$,
	(b)$\Delta_0/t_0=0.0$ and $\Delta^{SC}_0=0.3$, and
	(c)$\Delta_0/t_0=1.0$ and $\Delta^{SC}_0=0.0$.
	The angle is defined so that the value of $45$ is at the nodal point
	and the values of $0$ and $90$ are at the anti-nodal points.
    }
 \end{figure}

Figure \ref{fig_spA_along_fs} shows the spectral weight along the Fermi surface
at fixed $Q=(\pi,\pi)$.
%=====================================================================
% Data: Data0603_08/data1, data2, data3
% code: dd1_5.cc
%=====================================================================
In order to identify the contributions from the superconductivity correlation
and the orbital antiferromagnetic correlation,
we compare the result with or without each correlation.
One can see that
the structure around the superconducting gap node is mainly determined 
by the d-wave superconductivity correlation.
Meanwhile the antinode structure is mainly determined by
the orbital antiferromagnetic correlation.

\begin{figure}
   \begin{center}
    \includegraphics[width=3.4in]{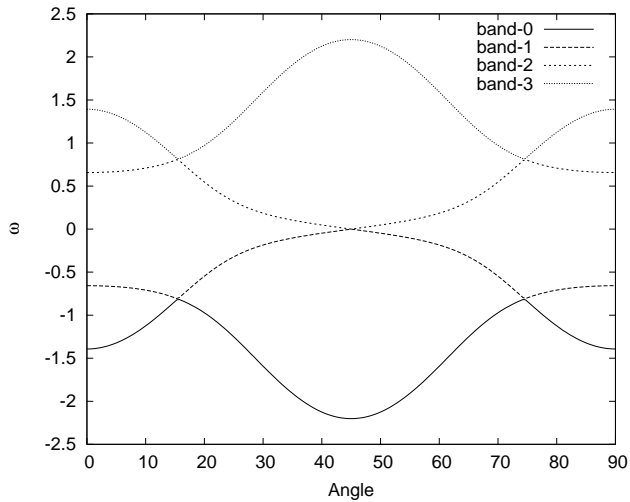}
   \end{center}
   \caption{ \label{fig_bare_band_dispersions}
	The energy dispersion along the underlying Fermi surface.
	The horizontal axis represents angle along the underlying Fermi surface.
    }
 \end{figure}

\begin{figure}
   \begin{center}
    \includegraphics[width=3.4in]{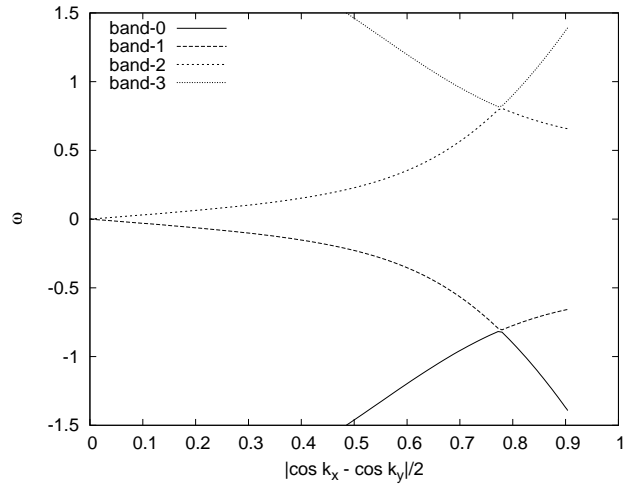}
   \end{center}
   \caption{ \label{fig_each_band_disp_cc}
	The dispersion energy versus $|\cos k_x - \cos k_y|/2$.
    }
 \end{figure}

Figure \ref{fig_bare_band_dispersions} shows the dispersion energies
along the underlying Fermi surface.
By comparing with Fig.\ref{fig_spA_along_fs},
we see that a part of the energy dispersions is invisible in
the spectral weight because of the coherence factor effect.
As shown in Fig. \ref{fig_each_band_disp_cc},
the energy dispersion around the node is well
described by d-wave gap function.
Around the anti-node, the energy dispersion is also well
described by d-wave symmetry.
However, this component is dominated by the orbital
antiferromagnetic correlation as suggested from Fig.\ref{fig_spA_along_fs}.
This analysis suggests that clear information about the 
superconducting gap is extracted near the node.
In fact, from the linear function fitting of the gap
near the node we find $\Delta_0^{SC}$ from Fig.\ref{fig_each_band_disp_cc}.
By contrast, the gap near the antinode contain both contributions
of the superconductivity correlation and the orbital antiferromagnetic
correlation.
It is hard to decompose these components even in our idealized model.

%=====================================================================
%=====================================================================

%=====================================================================
% Data: Data0604_08c/data1
% code: dd1_6.cc
%=====================================================================

%=====================================================================
% Data: Data0604_08b/
% code: dd1_5dos.cc
\begin{figure}
   \begin{center}
    \includegraphics[width=6cm]{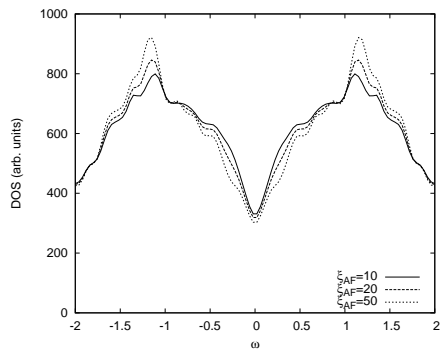}
   \end{center}
   \caption{ \label{fig_dos_Q_av}
	Energy versus density of states averaged over probability distribution
	with respect to $Q$ for different values of $\xi_{AF}$.
    }
 \end{figure}
%=====================================================================

%=====================================================================
% Data: Data0604_08b/
% code: dd1_5dos.cc
\begin{figure}
   \begin{center}
    \includegraphics[width=3.4in]{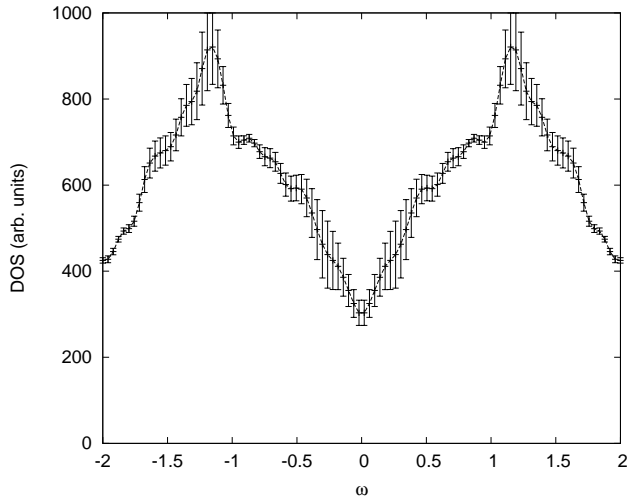}
   \end{center}
   \caption{ \label{fig_dos_Q_av_xiaf20}
	Energy versus density of states averaged over probability distribution	
	with respect to $Q$ at $\xi_{AF}=20$.
	Error bars are due to probability distribution of $Q$.
    }
 \end{figure}
%=====================================================================

Next we compute the density of states as follows
\be
D(\omega) = \sum_{k} 
%\sum_{\alpha=1,2,3,4}
\sum_{\alpha=1}^4
\delta \left( w-E_k^{(\alpha )} \right).
\label{eq_dos}
\ee
For numerical computations, we replace the delta function
with a Lorentzian.
The broadening factor is chosen as $0.10$.
In Fig. \ref{fig_dos_Q_av} we show $D(\omega )$ for different 
values of $\xi_{AF}$.
We can distinguish each band contribution.
The peaks around $\omega \simeq \pm 1$ are associated with
the orbital antiferromagnetic correlation.
While the shoulders around $\omega \simeq \pm 0.3$ are 
associated with the superconductivity correlation.
These features become sharp for large $\xi_{AF}$.
But the broadening effect due to $Q$ fluctuations
are similar as shown in Fig.\ref{fig_dos_Q_av_xiaf20}.
This result suggests that inhomogeneity of the relatively high energy
component is not associated with the antiferromagnetic correlation 
effect.

\begin{figure}
   \begin{center}
    \includegraphics[width=3.4in]{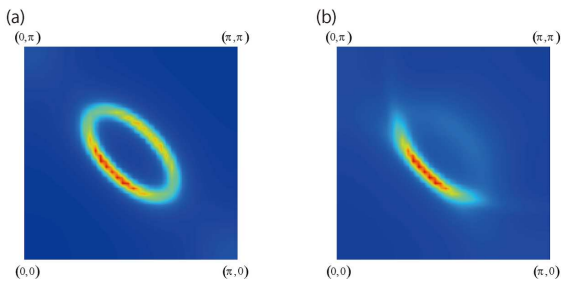}
   \end{center}
   \caption{ \label{fig_dos_af_effect}
	Density of states computed by Eq.(\ref{eq_dos}) 
	at $\omega=0.079$ in a quadrant of the Brillouin zone:
	(a)$\xi_{AF}=50$ and (b)$\xi_{AF}=5$.
	For both panels, we take $\Delta_0=1$ and $\Delta_{dSC}^0=0.3$.
    }
 \end{figure}

Figure \ref{fig_dos_af_effect} shows the intersection of $D(\omega )$
at $\omega=0.079$.
The same result is obtained at $\omega=-0.079$.
For large $\xi_{AF}$, we see some weight
outside of the reduced Brillouin zone.
By contrast, this weight disappears for short $\xi_{AF}$.
As a result, we see a banana shape
as seen in STM experiments.

%=====================================================================
% Data: Data0605_08b/data2(xiAF=5), data5(xiAF=50)/ dosq8.dat (w=0.0789)
% Ndiv=40, D_0=1.0, D_dSC^0=0.30, x=0.10, delta=0.10
% code: dd1_8.cc
% shell: sh_dd1_8
%=====================================================================

%\section{Summary}
%\label{sec_summar}
To summarize, we study the spectral weight and the density of states
in the presence of both of d-wave superconductivity correlation
and the orbital antiferromagnetic correlation with taking into account
the finite-range of the antiferromagnetic correlation effect.
Although these correlations are assumed to have the same d-wave symmetry,
it is demonstrated that these components are clearly distinguished
in the spectral weight and the density of states.
As for asymmetry of the density of states, 
we need to include impurity scattering effect 
as demonstrated in Ref.\cite{Ghosal2005}.

\begin{acknowledgments}
I would like to thank Prof. T. Tohyama for helpful discussions.
The numerical calculations were carried out in part 
on Altix3700 BX2 at YITP in Kyoto University.
\end{acknowledgments}

\bibliography{../references/tm_library}

\end{document}